\documentclass[11pt,a4paper]{article}
\usepackage{amsmath,amssymb}
\usepackage{epsfig,graphicx}
\usepackage[normalem]{ulem}
\usepackage[section]{placeins}
\newcommand{\be}{\begin{equation}}
\newcommand{\ee}{\end{equation}}
\newcommand{\bea}{\begin{eqnarray}}
\newcommand{\eea}{\end{eqnarray}}
\newcommand{\x}{{\bf x}}
\newcommand{\half}{\frac{1}{2}}
\begin{document}
\title{\vskip -70pt
\begin{flushright} {\normalsize {DAMTP-2014-49}} \end{flushright}
\vskip 30pt
\bf {States of Carbon-12 in the Skyrme Model}
}
\vskip 100pt
\author{\bf{\Large P.H.C. Lau}\footnote{P.H.C.Lau@damtp.cam.ac.uk} 
\quad {\bf {\Large and}} \quad 
\bf{\Large N.S. Manton}\footnote{N.S.Manton@damtp.cam.ac.uk}\\[30pt]
{\it Department of Applied Mathematics and Theoretical Physics,}\\[0pt]
{\it University of Cambridge, Wilberforce Road, 
Cambridge CB3 0WA, U.K.}\\[10pt]
}
\vskip 30pt
\maketitle
\vskip 30pt
\begin{abstract}
The Skyrme model has two Skyrmion solutions of baryon number $12$, with $D_{3h}$ 
and $D_{4h}$ symmetries. The first has an equilateral triangular shape
and the second an extended linear shape, analogous to the triangle 
and linear chain structures of three alpha particles. We recalculate the
moments of inertia of these Skyrmions, and deduce the energies and 
spins of their quantized rotational excitations. There is a good match 
with the ground-state band of Carbon-12, and with the recently established
rotational band of the Hoyle state. The ratio of the root mean square 
matter radii also matches the experimental value. 

\end{abstract}
\section{Introduction}

\begin{figure}[ht]
\centering
\includegraphics[width=4cm]{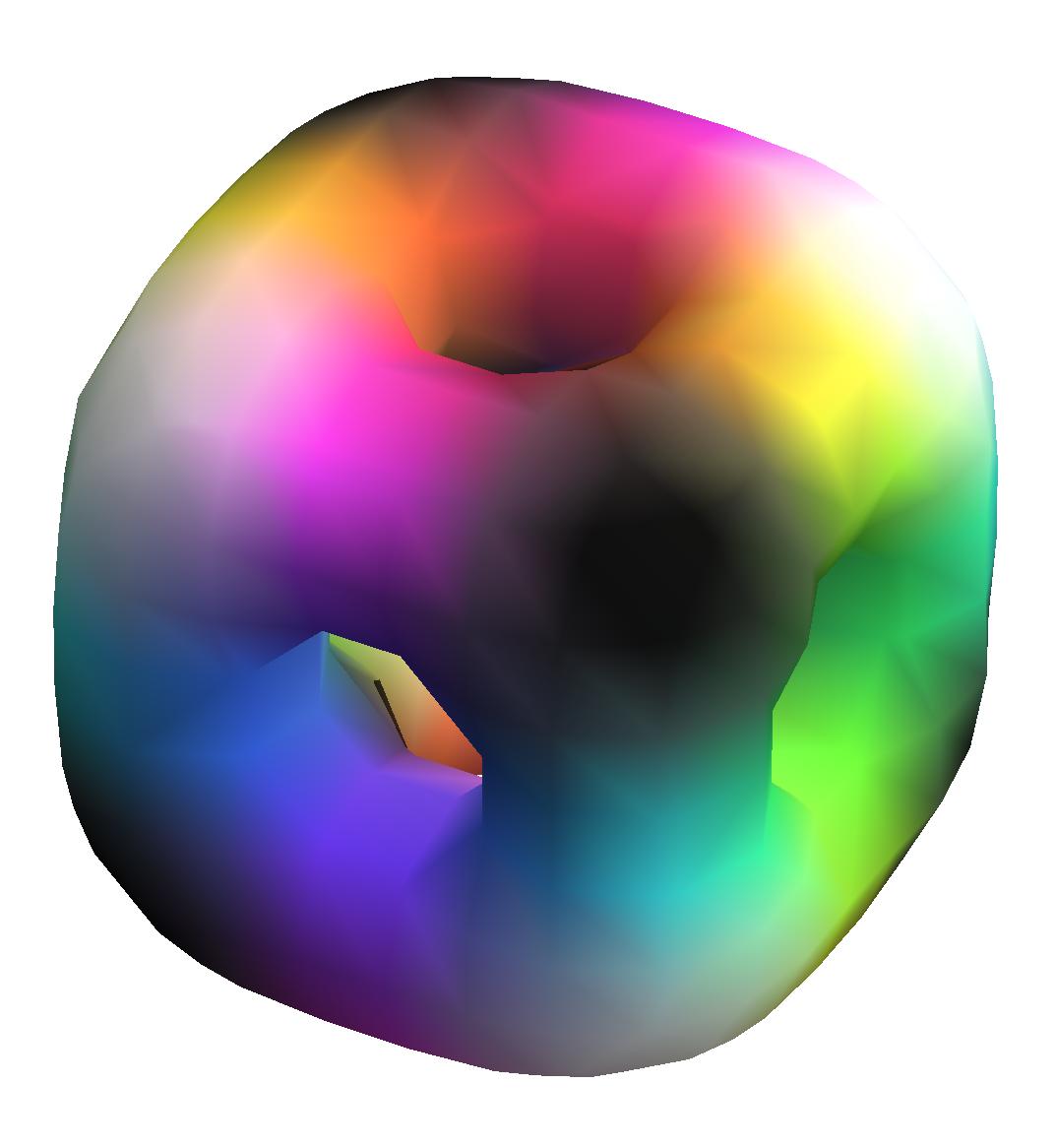}
\hspace{1cm}
\includegraphics[width=4cm]{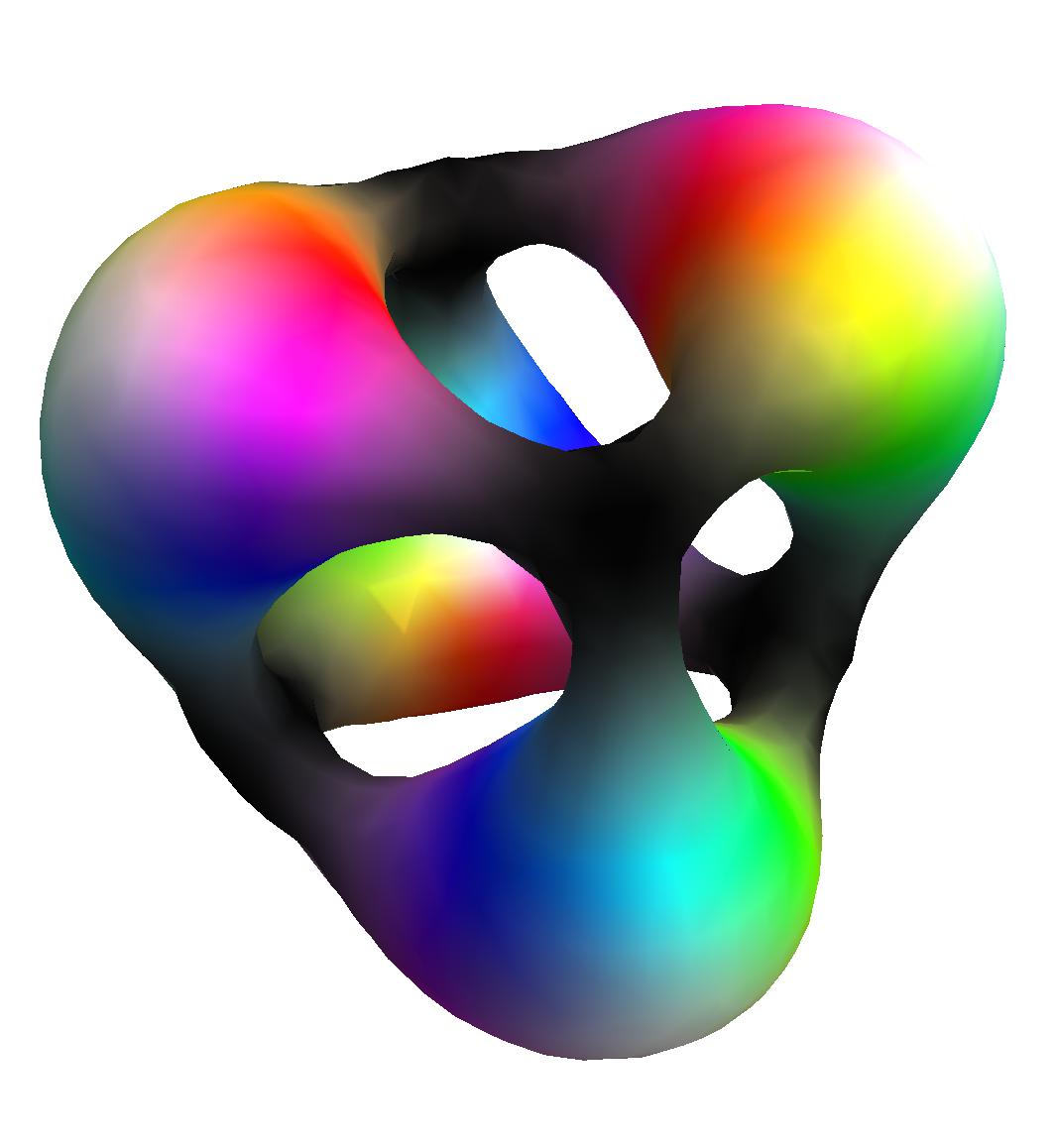}
\caption{$B=4$ Skyrmion (left) and deformed configuration (right). The figures show a constant baryon density surface and are coloured using P.O. Runge's colour sphere. The colours indicate the value of the unit pion field $\hat{\boldsymbol{\pi}}=\boldsymbol{\pi} / |\boldsymbol{\pi}|$ \cite{flm}.}
\label{fig:B_4}
\end{figure}

\begin{figure}[ht]
\centering
\includegraphics[width=7cm]{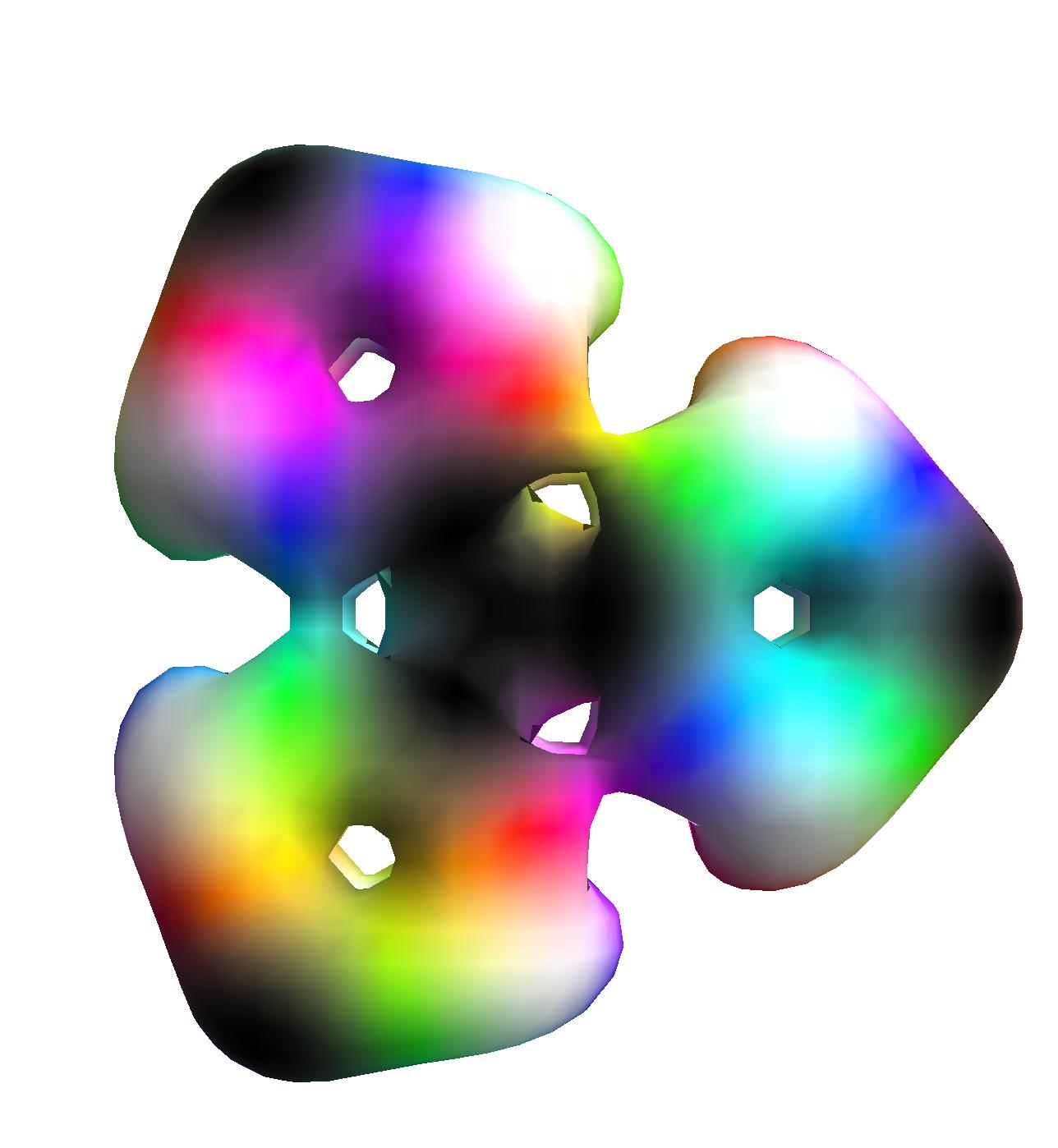}
\caption{$B=12$ Skyrmion with $D_{3h}$ symmetry.}
\label{fig:B_12_D3h}
\end{figure}

\begin{figure}[ht]
\centering
\includegraphics[trim = 10mm 100mm 5mm 50mm, clip, width=9cm]{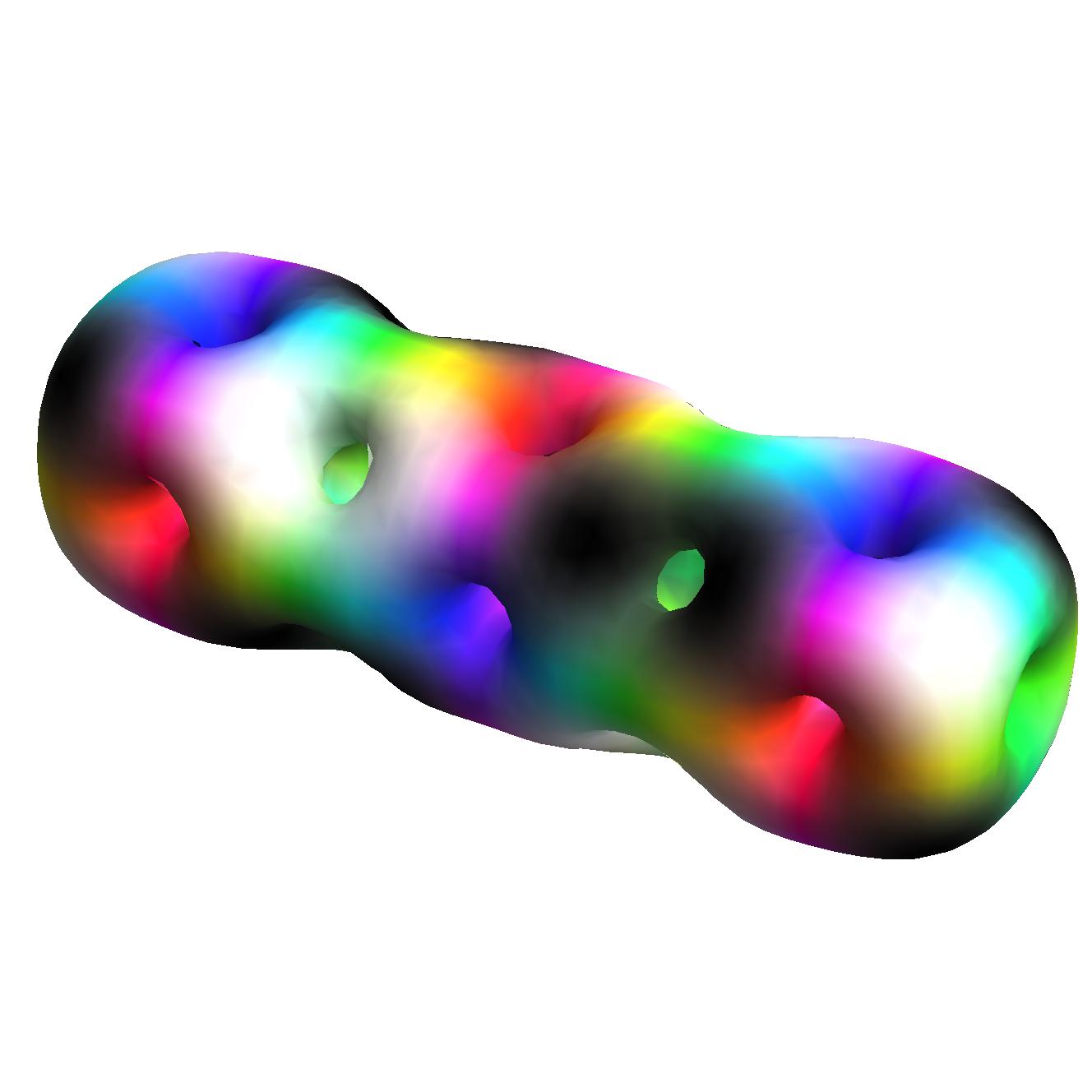}
\caption{$B=12$ Skyrmion with $D_{4h}$ symmetry.}
\label{fig:B_12_D4h}
\end{figure}

Static Skyrmion solutions of the Skyrme model with $B=12$, where $B$ is the 
baryon number, have been established for some time \cite{bms}. The 
solutions can be interpreted as three $B=4$ Skyrmions bound together. 
The $B=4$ Skyrmion, which has cubic symmetry \cite{btc,wal}, is 
particularly stable, and is a
building block for many further Skyrmions with $B$ a multiple of $4$.
The $B=4$ solution is illustrated in Fig. \ref{fig:B_4}, together with a
deformed configuration of slightly higher energy that is a 
tetrahedral arrangement of four $B=1$ Skyrmions, analogous to the
conventional picture of two protons and two neutrons in an alpha particle.
$B=12$ Skyrmions can be found numerically by allowing a symmetric 
arrangement of three $B=4$ Skyrmions to relax to a minimal energy 
solution. The initial arrangement may have the $D_{3h}$ symmetry of 
an equilateral triangle, or a straight chain structure, with $D_{4h}$ 
symmetry. In both cases, neighbouring cubes are oriented so as 
to strongly attract. The relaxed solutions retain the initial 
symmetries and have almost identical final energies. They are shown in
Figs. \ref{fig:B_12_D3h} and \ref{fig:B_12_D4h}. There is an energy barrier between these
solutions, because the cubes have to be moved apart a little, and 
rotated, to pass from one solution to the other. The chain solution is
one member of a family of chain solutions made from any number of
$B=4$ cubes \cite{hw}. The limiting, infinitely-long chain has a $45$-degree 
twist symmetry, and can be split into cubes in two independent ways. 

Details of the Skyrme model \cite{sky} have been reviewed elsewhere 
\cite{ManSut,BroRho,bmsw}. Briefly, it is a nonlinear field theory 
of pions where baryons are topological solitons. It is a type of 
nonlinear sigma model, with three pion fields
$\boldsymbol{\pi}(\x,t)$ combined into an SU(2)-valued Skyrme field
\be
U(\x,t)=\sigma (\x,t) {\bf 1}_2 +i\boldsymbol{\pi}(\x,t)\cdot 
\boldsymbol{\tau}\,.
\ee
The field $\sigma$ is not independent, because of the constraint  
$\sigma^2 + \boldsymbol{\pi}\cdot \boldsymbol{\pi}=1$. 

The Lagrangian in Skyrme units is
\be
L=\int \left\{ -\frac{1}{2}\,\hbox{Tr}\,(R_\mu R^\mu) 
+ \frac{1}{16}\,\hbox{Tr}\,([R_\mu,R_\nu][R^\mu,R^\nu]) 
+ m^2\,\hbox{Tr}\,(U - {\bf 1}_2) \right\} d^3 x \,,
\ee
where $R_\mu = (\partial_\mu U)U^{\dag}$, and $m$ is the dimensionless
pion mass. Physical units are obtained by fixing an energy and length scale
appropriate to nuclear physics. The value of $m$ has been suggested to be in the range of 0.68 to 1.15 \cite{bmsw}. This range of values gives a reasonable match to a range of nuclei. Here, we fix $m=1$. At the end of this letter, we will discuss the effect of changing $m$ to 0.7. The 
Lagrangian splits into a kinetic part, quadratic in time derivatives 
of $U$, and a static potential part. Skyrmions are minima of the 
potential energy and are labelled by their baryon number $B$, the topological 
degree of the field $U:{\mathbb{R}}^3 \rightarrow \rm{SU(2)}$ at a
given time, which is well-defined for fields satisfying the boundary condition 
$U \rightarrow {\bf 1}_2$ at spatial infinity. The baryon density is 
\be
{\cal B} = -\frac{1}{24\pi^2}\,
\epsilon_{ijk}\,\hbox{Tr}\, (R_i R_j R_k) \,,
\ee
and $B$ is the spatial integral of this. Here, $\mu,\nu$ are spacetime
indices and $i,j,k$ are spatial indices. 

To model nucleons and nuclei, one quantizes the Skyrmions as rigid
bodies \cite{anw}. Some vibrational excitations may also be included, but it is
not practicable to treat the Skyrme model as a quantum field
theory. In any case, one hopes that in the low energy regime of
nuclear physics, where free pion particles are not produced, a finite 
dimensional truncation of the Skyrme model
is sufficient. The Skyrme Lagrangian is invariant under
rotations in space (it has full Poincar\'e invariance but this is not
needed here), and isorotations $U \rightarrow A(t)UA(t)^\dag$, where 
$A \in \rm{SU(2)}$. Isorotations mix
the pion fields among themselves. A Skyrmion's mass is
its field potential energy, and it has a moment of inertia
tensor that arises from the field kinetic energy when the Skyrmion
rotates and isorotates. The formulae for the inertia tensors are 
now well known, but are rather complicated \cite{bc}. 

A quantized Skyrmion acquires spin and isospin. For even baryon numbers, 
these are integral. For the $D_{3h}$-symmetric $B=12$ Skyrmion, 
the rotational and isorotational motions are weakly 
coupled, but we ignore this, as it has a negligible effect on energy 
levels. There is no such coupling for the $D_{4h}$-symmetric Skyrmion. We
are mainly interested here in the states of isospin 0,
corresponding to Carbon-12, and just present the quantum Hamiltonian
for purely rotational motion. The quantized $D_{3h}$-symmetric Skyrmion
models the $0^+$ ground state of Carbon-12 and its rotational
excitations. The quantized $D_{4h}$-symmetric Skyrmion is identified
with the $0^+$ Hoyle state and its rotational
excitations \cite{hoy,whe,ff}. We discuss below whether this identification is reasonable.

\section{Quantizing the $B=12$ Skyrmions}

The two Skyrmions of interest, shown in Figs. \ref{fig:B_12_D3h} and \ref{fig:B_12_D4h},
have moment of inertia tensors of symmetric-top type, with distinct 
eigenvalues $V_{11} = V_{22}$, and $V_{33}$.  The 3-axis is the
$C_3$ or $C_4$ symmetry axis. $V_{33}$ is larger than $V_{11}$ for the
oblate, triangular solution, and smaller for the prolate, chain solution.   

In both cases, the quantum Hamiltonian for rotational motion is
\be
H =\frac{1}{2V_{11}}\mathbf{J}^2 
+ \left(\frac{1}{2V_{33}}-\frac{1}{2V_{11}}\right)K^2 \,,
\ee
where $\mathbf{J}$ is the quantum spin operator and $K$ its 
projection along the (body-fixed) 3-axis. The energy eigenvalues are simply
\be
E(J,k) = C\left\{\frac{1}{2V_{11}}J(J+1)
+ \left(\frac{1}{2V_{33}}-\frac{1}{2V_{11}}\right)k^2 \right\} \,,
\label{rotenergy}
\ee
where $J$ is the total spin label, and $k$ the eigenvalue of $K$, with
the projection chosen so that $0 \le k \le J$.
$C$ is a conversion factor from Skyrme units to physical units.

\begin{figure}[ht]
\centering
\includegraphics[trim = 0mm 15mm 0mm 20mm, clip, width=\textwidth]{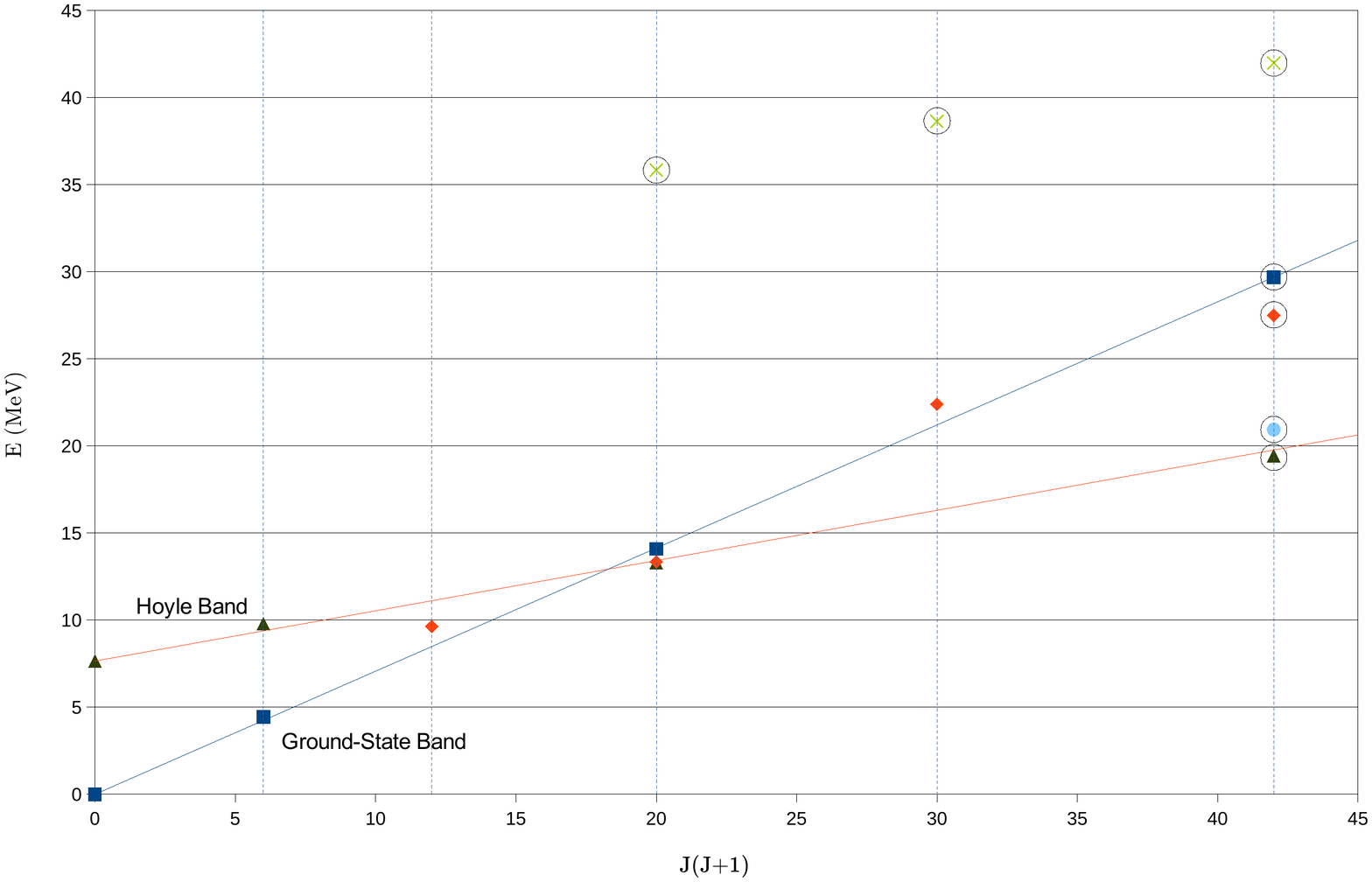}
\caption{Experimental states of Carbon-12. New states predicted by the Skyrme model are encircled. The symbols square, diamond, and circle denote the $k=0,3,6$ states of the ground-state band, and triangle and cross denote the $k=0,4$ states of the Hoyle band. The $0^+$, $2^+$, $4^+$ states have energies $0$, $4.4$, $14.1$ MeV for the ground-state band and $7.65$, $9.8$, and $13.3$ MeV for the Hoyle band \cite{whe}.}
\label{fig:C12_plot}
\end{figure}

The moments of inertia for the $D_{3h}$ triangular Skyrmion were calculated
before \cite{bmsw}, and we have confirmed them. They are 
$V_{11} = 5039$ and $V_{33} = 7689$. The numerical errors are of order 
$\pm 3$. The allowed rotational states for the $D_{3h}$ Skyrmion are exactly
the same as for an equilateral triangle of three identical, bosonic alpha 
particles. $k$ must be zero or a multiple of 3. Up to spin 6 the
allowed states have spin/parities 
$J^P = 0^+$, $2^+$, $3^-$, $4^-$, $4^+$, $5^-$, $6^+$, $6^-$, $6^+$, where the negative
parity states have $k=3$ and the second $6^+$ state has $k=6$. The
first two of these represent the ground state and first excited state of
Carbon-12. The $D_{3h}$ symmetry was fairly clear from the 
experimentally observed states up to spin 4, after the existence of the
$4^-$ state was clarified \cite{fre,kir}, and has been further confirmed by the 
recent observation of the $5^-$ state \cite{mar}. Higher energy states of
Carbon-12 with isospin 0 have been seen, but none are yet established
as having spin 6 \cite{whe}. In Fig. \ref{fig:C12_plot} we plot the energies and 
$J^P$ values of the observed 
states and the best fit of the formula (\ref{rotenergy}) to the $0^+$, $2^+$ and $4^+$ states of the ground-state band. $C$ is 
the fitted parameter, and has value $7130$ MeV. Because $V_{33}/V_{11} =
1.53$ in the Skyrme model, the states in the $k=3$ band 
are predicted to be about 2 MeV below those of the same spin in the 
$k=0$ band. The data are marginally consistent with
this. The three spin 6 states are characteristic of any 
model involving rigid rotations of an equilateral triangle, although 
their precise energy ratios depend on $V_{33}/V_{11}$.

\section{Hoyle states}

Because the ground state and Hoyle state of Carbon-12 are both $0^+$
states, their energies are simply the classical Skyrmion energies in our 
approach to the Skyrme model. Our best estimate is that the energies 
of the $D_{3h}$ and $D_{4h}$ Skyrmions are, respectively, $1816$ and
$1812$ in Skyrme units, with numerical errors which may be as large 
as $0.2\%$. The difference of these large numbers is very uncertain. As a 
result, we cannot confirm with any accuracy at 
all the 7.65 MeV energy difference between the Hoyle state and the 
Carbon-12 ground state, which is just $0.07\%$ of the
total rest energy of Carbon-12. Figs. \ref{fig:B_12_D3h} and \ref{fig:B_12_D4h} show that 
the $D_{4h}$ Skyrmion has two very strong bonds between the three $B=4$ cubes,
whereas the $D_{3h}$ Skyrmion has three rather weaker bonds, because
the cubes are differently oriented. As the $B=4$ Skyrmion has energy $613$, the classical bond energy is about
$10$, so it is not unreasonable that the ground and Hoyle states are
close in energy, but we cannot be more precise, even about their
ordering. Altogether, there is a problem in the Skyrme model
concerning nuclear binding energies, which we will not address here.

However, we can with some confidence study the slope of the rotational band 
based on the Hoyle state, and compare to the slope of the band based on the 
Carbon-12 ground state, because these depend on the moments 
of inertia. The moments of inertia for the $D_{4h}$ chain Skyrmion 
were only estimated previously \cite{wood}. The estimate used the parallel
axis theorem applied to three separated, undeformed $B=4$
Skyrmions. We have calculated these moments of inertia properly for 
the first time, and find $V_{11} = 12699$ and $V_{33} = 2106$. 
$V_{11}$, in particular, is rather greater than what was estimated 
(after extrapolating to $m=1$). For each of 
the $k=0$ bands there are observed $J^P = 0^+$, $2^+$, $4^+$ states. The experimental 
energies of the Hoyle state and its excitations are included in Fig. \ref{fig:C12_plot}. 
The Skyrme model prediction for the ratio of the slopes is just the 
ratio of the $V_{11}$ values for the $D_{4h}$ and $D_{3h}$ Skyrmions, 
which is $12699/5039 = 2.52$. The
dimensional conversion factor $C$ cancels. We estimate the ratio of the experimental slopes from the best linear fit to the $0^+$, $2^+$ and $4^+$ states in Fig. \ref{fig:C12_plot}. For the ground-state band the best fit slope is $0.707$ MeV, and for the Hoyle band it is $0.289$ MeV. The
ratio is $2.45$, agreeing with the prediction. 

The ratio $2.52$ reflects the extended structure and separation of the $B=4$
Skyrmion subunits. For three ideal point alpha particles with a fixed bond
length separating them, arranged as an equilateral triangle or as a linear
chain, the ratio of the $V_{11}$ values is $4$. In the Skyrme model the 
ratio is smaller, partly because the bond is less tight in the $D_{3h}$
Skyrmion, but mainly because of the extended form of the $B=4$ cubes.   

In total, we predict five states of spin 6, three in the Carbon-12
ground-state band (with $k=0,3,6$) and two in the Hoyle band (with $k=0, 4$). Only the $k=3$ state has negative parity. The predicted
energies are, respectively, $E= 29.7$, $27.5$, $20.9$, $19.4$ and $42.0$ MeV. The Hoyle state
excitations with $k=4$, starting with $4^+$ and $5^+$ states, are of
considerably higher energy because of the strongly prolate nature of 
the $D_{4h}$ Skyrmion.

\section{Matter Radii and Vibrational Modes}

The root mean square matter radius $\langle r^2 \rangle^{\half}$ 
provides a further test of the Skyrme model. For each Skyrmion we 
can calculate
\be
\langle r^2 \rangle^{\half} = \left(\frac{\int \rho(\x) r^2 \, d^3x}
{\int \rho(\x) \, d^3x} \right)^{\half} \,,
\ee
where $\rho(\x)$ is the static energy density, interpreted as a mass
density, and $r = |\x|$. We find that the $D_{3h}$ and $D_{4h}$
Skyrmions have $\langle r^2 \rangle^{\half}$ values $2.29$ and $2.80$, 
respectively, in Skyrme units. The ratio is $1.22$. The experimental 
matter radius for the ground state is $2.43$ fermi \cite{tan} and for
the Hoyle state it is inferred to be $2.89$ fermi \cite{dan}. Their ratio
is $1.19$. Again the prediction of the Skyrme model is very good.

Replacing $\rho$ by the baryon density ${\cal B}$ gives radii $1\%$ 
smaller, as the baryon density has a more compact tail than the energy
density. These radii are the predictions for the root mean 
square charge radii of the Carbon-12 ground state and Hoyle state, because, 
for isospin 0 states, the charge density is half the baryon density 
in the Skyrme model. The ratio is still $1.22$. Including Coulomb
energy effects may change the ratio a little.

In the point alpha particle model the ratio of matter radii is
$\sqrt{2}$, and other models predict larger ratios \cite{che}. The smaller
experimental ratio has been obtained in lattice calculations of 12-nucleon states, using interactions derived from chiral effective field theory, where the Hoyle 
state is found to be an obtuse 
triangular structure \cite{epl}. The Skyrme model seems to favour a straight chain of alpha particles. The straight chain, with its large $D_{4h}$ symmetry 
group, admits far fewer low-lying rotational states than an obtuse 
triangle, just the $J^P = 0^+$, $2^+$, $4^+$ states for which there is 
clear experimental evidence. So, arguing from the Skyrme model, we 
expect that Hoyle state excitations with $J^P = 3^-$ and $4^-$, suggested
in \cite{mar}, will not be seen.

Vibrational spectra of the $B=12$ Skyrmions have not been studied, but
the lowest vibrational modes of the $D_{3h}$ Skyrmion are likely to be
the degenerate pair of triangle-deforming modes. These should give rise to
the observed $J^P = 1^-$, $2^-$ states of Carbon-12, but we do not have
an estimate for their energies. The potential energy landscape is rather flat
for this vibrational excitation, particularly in the direction where the
triangle becomes obtuse, so the vibrations may not be close to
harmonic. The first excited state of this vibrational
mode could be quite similar to a state based on the obtuse 
triangle in a rotational excitation. The $D_{3h}$-symmetric breathing 
vibrational mode will give rise to a $0^+$ state, probably of higher 
energy \cite{fei}, and not to be identified as the Hoyle state.   

\section{Effect of changing $m$}

Properties of Skyrmions such as the moments of inertia and energy depend on the value of $m$. It is reasonable to ask how sensitive our results are to the value of $m$. We repeated our analysis using a value of $m=0.7$, the choice preferred in \cite{bmsw}. In Skyrme units, a Skyrmion gets larger and less massive when $m$ is reduced. But recalibration of the length and energy scale compensates for this, and the nuclear properties change rather little. We calibrated the Skyrme parameters using the rest mass and the slope of the ground-state band of Carbon-12, so these quantities are the same for $m=0.7$ as for $m=1$. For $m=1$ our calibration gives a pion mass of 210 MeV; for $m=0.7$ it gives 162 MeV, which is closer to the experimental value 138 MeV.

In Skyrme units, when $m$ is reduced from 1 to 0.7, the matter radius of the $D_{3h}$ Skyrmion increases by 10\% while the matter radius of the $D_{4h}$ Skyrmion increases by 7\%. The ratio of these radii decreases from 1.22 to 1.19, a decrease of only 3\%. The new ratio matches exactly with the experiments. In Skyrme units, the moments of inertia of the Skyrmions also increase because of the increase in size. The moment of inertia $V_{11}$ of the $D_{3h}$ Skyrmion increases by 15\% while the moment of inertia $V_{11}$ of the $D_{4h}$ Skyrmion increases by 7\%. This gives a ratio of these moments of inertia, and hence of the slopes of the rotational bands, of 2.34 which is 7\% lower than before. This ratio is about 4\% lower than the experimental value, whereas the ratio for $m=1$ is about 3\% higher.

\section{Conclusions}

We have seen that the Skyrme model gives a quantitative
understanding of the spectrum of rotational excitations of Carbon-12,
including the excitations of the Hoyle state. The ground state 
is interpreted as an equilateral triangle of $B=4$ Skyrmions, modelling 
three alpha particles, and the Hoyle state is a linear chain of
these. Calculations of the moments of inertia and the ratio of the root mean square 
matter radii support the linear chain interpretation. It would be good 
to calculate electromagnetic transition strengths between the states, using the
Skyrme model, but the technology for this needs to be developed.

We end by recalling that within the Skyrme model, isospin excitations
are treated in the same way as spin excitations. The allowed isospin
and spin combinations are determined by the topological
Finkelstein--Rubinstein constraints \cite{fr}, which encode the Pauli principle
for nucleons within the Skyrme model, combined with symmetry considerations \cite{kru}. Calculations for the
isospin 1 and isospin 2 excitations of the $D_{3h}$ Skyrmion have been
performed before \cite{bmsw}. The four lowest lying states with isospin 1
have spins $J^P = 1^+$, $1^+$, $2^-$, $2^+$, and have energies close to the
lowest observed states of Boron-12, Nitrogen-12, and the isospin 
1 states of Carbon-12. They have a total energy separation of about 
2 MeV, which is correct, but the $2^+$, $2^-$ states are in the wrong
order, and two $1^+$ states have not yet been resolved 
experimentally. 

The $D_{4h}$ Skyrmion also has isospin excitations. Formulae 
for the quantum Hamiltonian and its energies were found by Wood 
\cite{wood}, but now that we have calculated the inertia tensors accurately, these
energies should be recalculated. In addition to the values $V_{11}$
and $V_{33}$ given above, we find for $m=1$ isospin moments of inertia $U_{11} = 437$,
$U_{22} = 449$ and $U_{33} = 472$, and for $m=0.7$, $U_{11} = 496$,
$U_{22} = 529$ and $U_{33} = 537$. There are no off-diagonal terms 
or spin/isospin cross terms here. A key prediction is that the lowest
isospin 1 excitation of the Hoyle state has $J^P = 0^-$, with energy 
a few MeV above the four isospin 1 states of the $D_{3h}$ Skyrmion 
mentioned above.

\section*{Acknowledgments}

We are grateful to Physics World (August 2014) for drawing our
attention to reference \cite{mar}. We are also grateful to Dankrad 
Feist for his work developing the codes to compute Skyrmions and their
properties. This work has been partially supported by STFC grant ST/J000434/1. P.H.C. Lau is supported by Trinity College, Cambridge.

\end{document}